\documentclass[aps,prl,reprint]{revtex4-1}
\usepackage{blindtext}
\usepackage{graphics}
\usepackage{subfigure}
\usepackage{epsfig}
\usepackage{epsf,epic}
\usepackage{color}
\usepackage{marvosym }
\usepackage{subfigure}
\usepackage{amsmath}
\usepackage{amssymb}
\usepackage{amsfonts}
\usepackage{wrapfig}
\usepackage{pstricks}
\usepackage{multirow}
\usepackage{pst-node}
\usepackage{bm}
\usepackage{dcolumn}
\newcommand{\etal}{\textit{et al.\ }}

\begin{document}
\title{Understanding the crystallographic phase relations
  in alkali-trihalogeno-germanates 
  in terms of ferroelectric or antiferroelectric arrangements
  of the tetrahedral GeX$_3$ units}
\author{Santosh Kumar Radha}
\email{santosh.kumar@case.edu}
\author{Walter R. L. Lambrecht}
\email{walter.lambrecht@case.edu}
\affiliation{Department of Physics, Case Western Reserve University, 10900 Euclid Avenue, Cleveland, OH 44106-7079, USA}
\begin{abstract}
 {\bf The alkali-trihalogeno-germanates AGeX$_3$ with A a large single positive ion
such as Rb, Cs, or organic radicals such as methyl ammonium (MA), and X a
halogen (I, Br, Cl, F) along with the corresponding
stannates (ASnX$_3$) and plumbates (APbX$_3$) exhibit a large variety of crystal structures, some of which are  of the perovskite type. 
These materials, better known  as ``halide perovskites'' 
have recently gained worldwide attention as promising photovoltaic
and more broadly opto-electronic materials. 
But their stability problems relative to the 
non-perovskite phases is a major issue.  Here we show that the
phase relations in these materials can be understood in terms of
the relative orientation of the GeX$_3$ tetrahedral units, which
is ferroelectric in the perovskite phase but antiferroelectric
in the competing phases. This suggests that an applied electric field
could be used to stabilize the desired phases and trigger a phase transition
between two phases of the material with widely different optical and electronic
properties.}
\end{abstract}
\maketitle

{\bf Introduction:} The hybrid halide perovskites have become a major new class of
photovoltaic materials with 
record efficiencies being  established in a very short development time.
\cite{snaith01,Park15} At the same time these materials have generated
a large interest from the scientific community to understand
what underlies their success as solar cell materials. In the
process it has become evident that 
these materials show a rather  unusual combination of ionic properties,
such as large Born effective charges,
and large $\varepsilon_0/\varepsilon_\infty$ with (typical for  covalent
materials)
small band gaps and exciton binding energy and a flexible inorganic
network with interesting dynamics.\cite{Lingyi13,Yaffe17}

Most of these favorable properties for opto-electronics rely on the electronic
band structure, which shows small direct gaps (for halides) close to the optimum
range (1.2-1.8 eV) for single or tandem solar cells
and relatively small carrier effective masses, in particular for holes.\cite{Lingyi13}
However, a main difficulty with these materials is their thermodynamic
stability, not only under environmental effects 
such as humidity or light exposure but intrinsically due to 
the  existence of other competing
phases which do not share these favorable features in the band
structures. In fact, these other phases may well be an intermediate step in the
decomposition process of the material
to the AX and (Ge,Sn,Pb)X$_2$ reaction products. 
It is thus important to understand the relative stability
and relations between these crystallographic phases, how they influence
the band structures and how their trends depend on the chemical substitution
space.

Some of the relations within the perovskite type of structures
are already well understood. Specifically, we showed recently that
both the Sn and Pb based compounds in this family prefer octahedral rotations
related to the Goldschmidt tolerance factor $t<1$
while the Ge and Si based ones show a ferroelectric
off-centering of the central IV-atom leading to a rhombohedrally distorted
perovskite.\cite{Santosh18} The rotational distortions in CsSnX$_3$
were studied both experimentally\cite{Inchungjacs} and computationally\cite{Lingyi14} and are well known from oxide perovskites. Similar rotated
octahedron phases also occur for the plumbates but are further complicated
in the hybrid organic ones by the symmetry breaking of the organic ion.
However, the relation between the perovskite and the non-perovskite phases
such as the yellow phase\cite{Lingyi17} in CsSnI$_3$ or the monoclinic phase\cite{Lingyi15} of CsSnCl$_3$ are not yet understood. These phases are usually
described in terms of edge-sharing rather than corner-sharing octahedra.

Because in the stannates the yellow phase has a higher density and
the rotations also are clearly driven by the need to make the space for the
alkali ion tighter, we proposed in previous work that to avoid
the non-perovskite phases one needs to make the size of the IV-X network
smaller relative to the A filler cation. First this already explains
why larger organic ions are preferred to Cs in the plumbates but also
guides the way to how to develop lead-free alternatives. 
This naturally led us to explore the
Ge and Si  based materials.\cite{Lingyi16,Lingyi16-csgevib,Radha18}
In fact, the CsGeX$_3$ family is found  not to exhibit octahedron
rotational distortions but a ferroelectric rhombohedral phase \cite{Thiele87} at low temperature and a cubic perovskite phase at high temperature
and was shown to have a band structure maintaining the favorable features
of the perovskite structure. 
However the situation is different for RbGeX$_3$, where the RbGeCl$_3$
shows a monoclinic $P2_1/m$ phase\cite{Messer78}
while RbGeBr$_3$ \cite{Thiele88} and RbGeI$_3$ \cite{Thiele89} show
a low temperature orthorhombic $Pn2_1a$ phase, and in the case of
RbGeI$_3$ another orthorhombic $P2_12_12_1$, besides the rhombohedral $R3m$ and
cubic $Pm3m$ phases. 
\begin{figure}
\includegraphics[width=7.5cm]{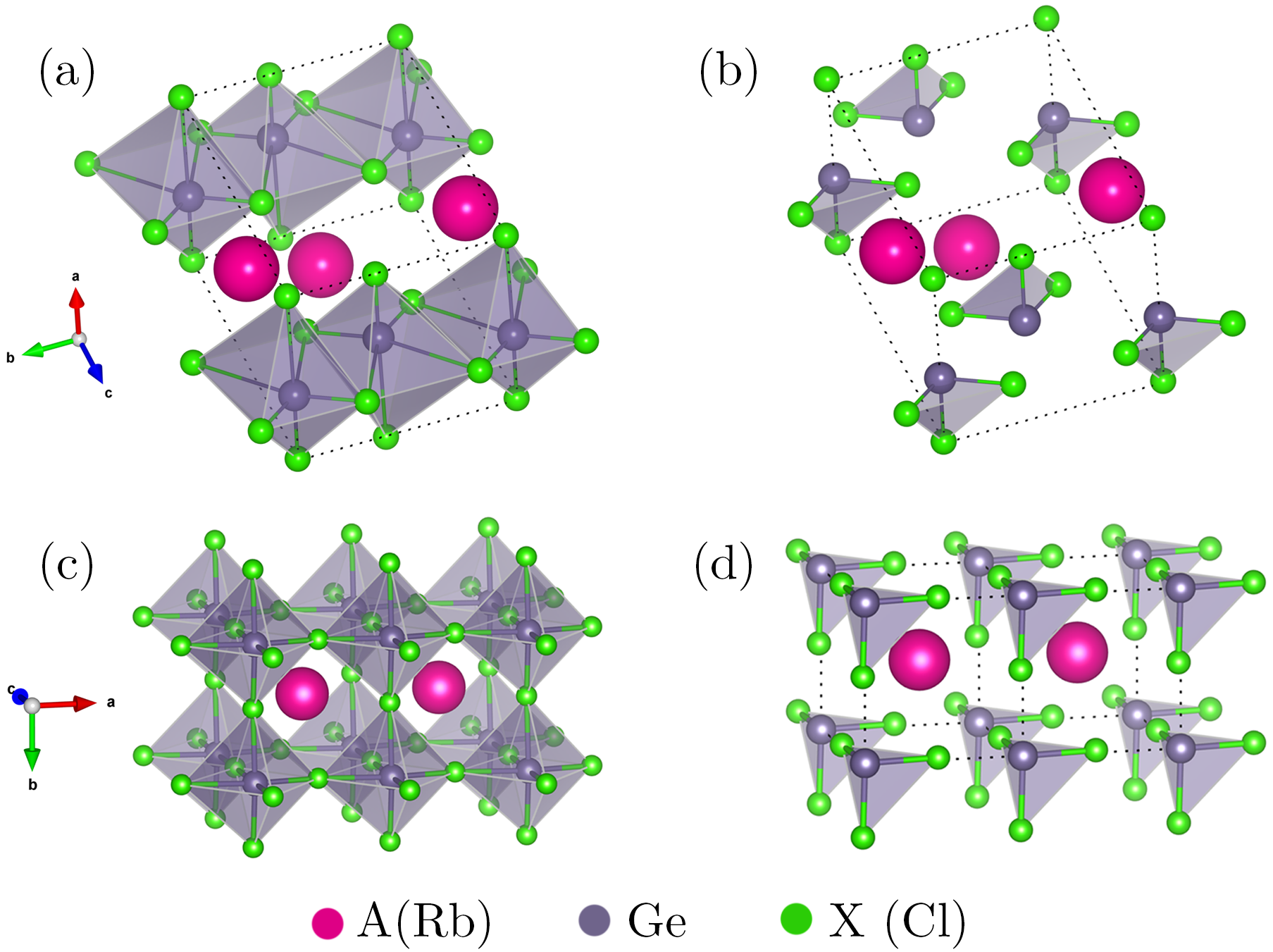}
      \caption{Monoclinic (a,b) and Rhomobohedral (c,d) structure viewed either
        as octahedral coordination (a,c) or tetrahedral (b,d)
        coordination of Ge.\label{figoctet}} 
\end{figure}

In this paper we show
that the relation between these phases can be better understood in terms
GeX$_3$ tetrahedral units. We focus on the relation and relative
stability between the monoclinic and rhombohedral phase. Even within
the rhombohedral phase, the Ge off-centering in its octahedron
results naturally in Ge making stronger bonds with three of its six halogen
neighbors. This results from the lone-pair chemistry of divalent Ge(II)
\cite{Santosh18} and makes  GeX$_3$ a natural
motif in terms of which to describe the structures. These units clearly
carry an electric dipole and as we will show below the monoclinic
phase is then simply an antiferroelectric arrangement of these dipoles while
the rhombohedral one is a ferroelectric one. We show using
first-principles calculations that the monoclinic phase is favored
for the RbGeX$_3$ while the rhombohedral one is favored for the CsGeX$_3$.
We then introduce an effective spin-model to explain why the electrostatic
dipole interactions favor the antiferroelectric alignment while
the ferroelectric alignment is stabilized by the additional bond formation.
The trends in the total energy differences in the family can be understood
on the basis of that model. Most importantly, the key role of electrostic
dipole interactions in the stability suggests that an applied electric
field can be used to trigger the  phase transition between them. 
Finally, we address how these different structures
affect the electronic band structure.  

{\bf Results:} In Fig. \ref{figoctet} we show the monoclinic
$P2_1/m$ and the rhombohedral perovskite $R3m$  crystal structures
of RbGeCl$_3$ viewed either with octahedral coordination of Ge 
or tetrahedral coordination. The perovskite structure is shown
in a doubled cell for easy comparison to the monoclinic structure.
We can see that while in the perovskite structure, the octahedra are
corner-sharing and form a 3D network, in the monoclinic structure, the
octahedra form 1D chains with edge-sharing octahedra.  The octahedra
are severely distorted and include three bonds of 2.35 \AA, one of 3.56 \AA\
and two of 4.06 \AA. In the perovskite phase, the three short bonds are 2.43 \AA\ and the three long ones are 2.90 \AA. It is clear that the GeX$_3$ tetrahedral units are pointing in opposite directions
in the monoclinic structure, while they point in the same direction
in the perovskite structure. In Supplemental Material (SM)\cite{SupM}
we show in more detail that one can simply rotate one of the 
GeX$_3$ units of the monoclinic structure around the {\bf b}-axis
about its center and hence arrive at the perovskite structure after
letting the structure relax by straightening out the network of
connected Ge-Cl bonds and letting the Rb find its optimum position. 
The Ge atoms form a simple cubic lattice connected via Cl
in cubic perovskite and this same lattice persists in the rhombohedral and
monoclinic structures in a distorted form but with the same topology.

  \begin{table}[t]
    \caption{Relative structural stability of monoclinic {\sl vs.} rhombohedral
      structure for CsGeX$_3$ and RbGeX$_3$.
      $V_{ratio}=V_{mono}/2V_{rhombo}$ is the volume ratio of the respective
      volumes per formula-unit, $\delta E =E_{rhombo}-E_{mono}$ the total energy
      difference       per formula unit. 
	\label{tabtote}}
	\centering
	\begin{tabular}{@{}llll@{}}
		\toprule
		Crystal & Stable Structure  & $ V_{ratio} $ & $ \delta E $(eV)  \\ \cline{1-4}
		&            &              &         \\
		RbGeCl$_3 $ & Monoclinic & 1.133        & 0.095  \\
		 CsGeCl$_3  $& Rhombohedral     & 1.094        & -0.034 \\ \cline{1-4}
		&            &              &         \\
		 RbGeBr$_3 $ & Monoclinic & 1.131        & 0.043   \\
		 CsGeBr$_3 $ & Rhombohedral       & 1.097        & -0.105  \\  \cline{1-4}
		&            &              &         \\
		 RbGeI$_3 $  & Monoclinic & 1.149        & 0.054 \\
		 CsGeI$_3 $  & Rhombohedral   & 1.093        & -0.169 \\ \cline{1-4}
	\end{tabular}
  \end{table}

Next, we examine the results of first-principle calculations of the
relative total energies of these two structures. 
Both structures were fully relaxed and the details of the calculations, 
the crystal structure and Wyckoff positions are given in SM.\cite{SupM}
The monoclinic structure is seen to be the lower energy
one for all Rb cases while the perovskite structure has lower energy
for the Cs compounds.  The volume per formula unit (f.u.) is always larger
in the monoclinic structure than in the rhombohedral structure
for the same compound.  However the volume ratio is systematically
larger for the Rb cases than the Cs cases.

To address the relative stability of the two structures, we first consider
the dipole electrostatic interactions. 
In a simple cubic lattice the net interaction energy of
dipoles pointing in the [111] (or $[\bar{1}\bar{1}\bar{1}]$) directions
beyond nearest neighbors is significantly smaller than that of the
nearest neighbors.
Thus, we can map the electrostatic problem
to that of dipoles or classical up-down spins in a fixed direction on
a simple cubic lattice and with nearest neighbor interactions only. 
The key point however is that the spins
have an anisotropic interaction. 
Writing the dipole part of the Hamiltonian as
\begin{equation}
  H_{dip}=J_\parallel\sum_{\langle ij\rangle} S_{i\parallel}S_{j\parallel}
  +J_\perp\sum_{\langle ij\rangle} {\bf S}_{i\perp}\cdot{\bf S}_{j\perp}
\end{equation}
with the sums over nearest neighbor pairs, and $S_{(i,j)\parallel}$  means
a spin on a pair of sites with connection vector parallel to the spin
and $S_{(i,j)\perp}$ spins on sites with connection vector perpendicular to the spin. Taking $J_{\parallel}=V^\parallel_{\uparrow\uparrow}-V^\parallel_{\uparrow\downarrow}$ and $J_\perp=V^\perp_{\uparrow\uparrow}-V^\perp_{\uparrow\downarrow}$,
with
$V_{ij}=[{\bf p}_i\cdot{\bf p}_j-3({\bf p}_i\cdot\hat{\bf r}_{ij})({\bf p}_\cdot\hat{\bf r}_{ij})]/r_{ij}^3$ the classical dipole interation
we obtain  
\begin{equation}
  J_\parallel=-4\frac{q^2d^2}{a^3}, \qquad
  J_\perp=2\frac{q^2d^2}{a^3} \label{eqJdip}
\end{equation}
which will favor parallel alignment for neighbors in the direction
parallel to the spin and anti-parallel alignment for neighbors
in the direction perpendicular to the direction of the spin.  
Here $d$ is the distance between the X$_3$ plane and the Ge in the GeX$_3$
tetrahedron and $q$ the effective charge of the dipole in this molecule
and $a$ is the cubic lattice constant. 

Spin and dipole interactions on a simple cubic lattice were studied
by Luttinger \etal\cite{Luttinger51,Luttinger46} and the anisotropic
exchange case was discussed by Belorizky \etal\cite{Belorizky} These
studies show that the preferred arrangement of spins is essentially
the one observed in the monoclinic structure,
namely, neighboring spins are parallel in the
direction in which the dipole points and antiparallel in the perpendicular directions. 
All we need to favor
this arrangement is  $J_\parallel<0$ and $J_\perp>0$ which is obeyed by Eq. \ref{eqJdip}.
The conclusion from this is that the electrostatics of dipoles
by itself would always favor the antiferroelectric alignment.
In order to explain the possible stability of the ferroelectric alignment
we then need to take into account
that for this arrangement of the GeX$_3$ additional
bonds can be formed by the Ge with the slightly further away halogens 
in the neighboring GeX$_3$ units, thereby in fact restoring
the 3D corner-sharing octahedral environment.  This could
be described  by adding a term to the spin Hamiltonian of the form 
$H_{bond}=-U\sum_{\langle ij\rangle}\Theta(S_{i\parallel}S_{j\parallel})$
with $\Theta$ the step function.

\begin{figure}[!h]
\includegraphics[width=6cm]{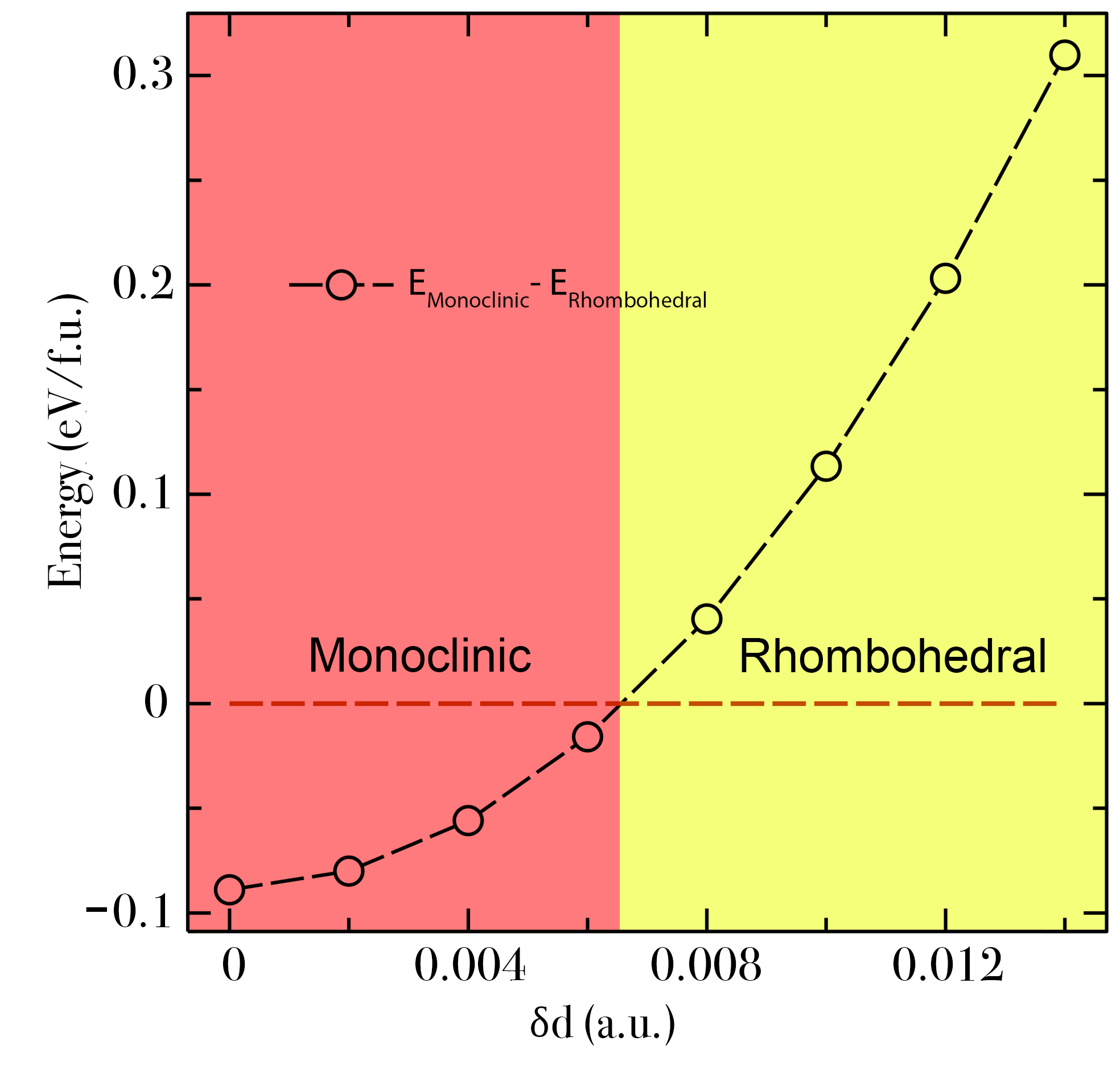}
\caption{Total energy difference between monoclinic and perovskite
  as function of the dipole strength varied by changing $\delta d$
  in the GeCl$_3$ molecular unit. 
    \label{figdipvary} 
  }
\end{figure}

The key point is that the dipole energy $H_{dip}$ and
the bond energy $H_{bond}$ must be of similar magnitude to have a meaningful
competition. This is apparently the case from our first-principles results.
To further illustrate this we calculated the energy difference between
the two structures for RbGeCl$_3$
while artificially changing the dipole strength.
We can do this by adjusting the parameter $d$ in the GeX$_3$ unit. 
Positive $\delta d$ (the deviation from the optimum energy $d$)
implies a smaller dipole and hence a reduction of the electrostatic
stabilization. We can see in Fig. \ref{figdipvary} that for $\delta d>\delta d_{crit}\approx0.006$ the perovskite phase becomes favored.

Furthermore this model can now be used qualitatively to explain the
trends in the total energy results.  The larger Cs atom
leads to an overall larger lattice constant $a$ and hence a reduction of the dipole term compared to the Rb case.  If we assume that the bond
formation energy depends less strongly on $a$
then this explains the basic finding why
the CsGeX$_3$ favor the perovskite while the RbGeX$_3$ prefer the
monoclinic structure.  Within the Cs family the dipole term becomes stronger
as we go from less ionic I to more ionic Br and Cl. Indeed we see that
for the Cs cases, the $\delta E=E_{rhombo}-E_{mono}$
in Table \ref{tabtote}
is less negative for Cl than Br than I, indicating the increasing importance
of the electrostatic dipole stabilization term. For the Rb cases, also
the highest $\delta E>0$ occurs for the most ionic Cl case.  The
$\delta E$ however is lower for the Br than the I case.
In fact, in the Br case, the reported low-temperature ($T<93^\circ$ C)
structure is not monoclinic but another
(layered) antiferroelectric arrangement of the GeX$_3$ dipoles
which already corresponds to a 3D network of corner-sharing octahedra
but which now combines octahedral rotations with Ge off-centered
distorted octahedra (see SM\cite{SupM}) and was calculated to have a 5.5 meV/f.u
lower energy than the monoclinic structure. 
The $P2_12_12_1$  structure of RbGeI$_3$  (for $T<181^\circ$ C) on the other hand
shows still another arrangement of the GeI$_3$ tetrahedra,
which when viewed as octahedra shows 1D chains of Ge$_2$I$_6$
units comprised of edge-sharing octahedra. This structure is a  closely related
variant of the $Pnma$ structure of CsSnI$_3$, the so-called yellow phase,
and is here found to have 10.7 meV/f.u. lower energy than the
monoclinic structure. 
In a separate paper we plan to  present a more elaborate model in
which we explore the energy landscape as function of fully rotating
the rigid GeX$_3$ tetrahedra in all possible directions
combined with an  energy term that counts
the number of Ge-X bonds being formed as the units rotate.  This model indeed
suggests that other (metastable) structures
with intermediate relative orientations
of the GeX$_3$ relative to each other exist. The model presented here in principle also applies to the stannates and plumbates although the
lone-pair related distortion in these cases is less pronounced  and
the $s^2$ electrons behave more as an inert-pair.

\begin{figure}
  \includegraphics[width=6.5cm]{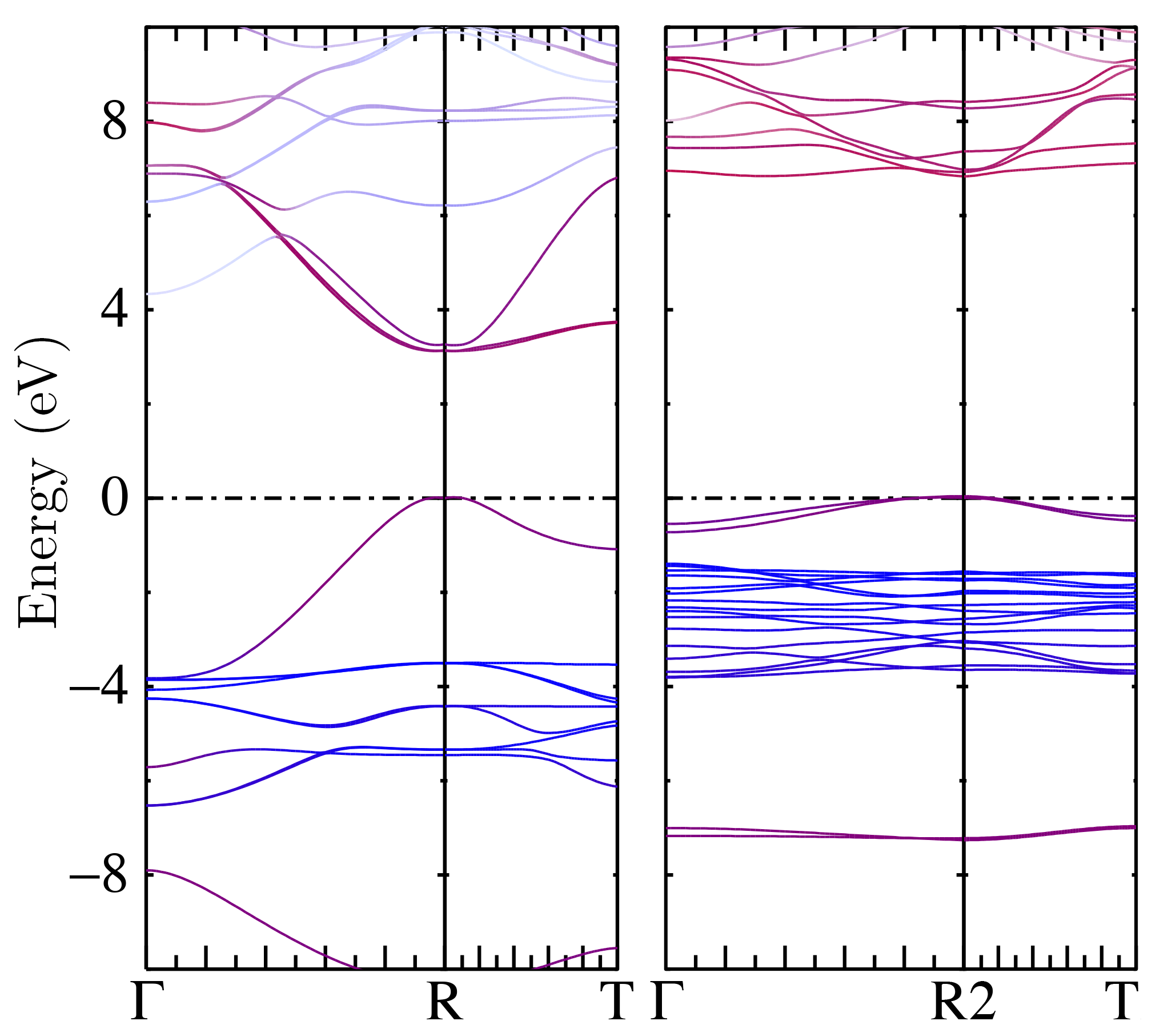}
  \caption{Band structure of RbGeCl$_3$ in perovskite
    and monoclinic structure. The blue color indicates Cl and
    the red Ge, the hybridized bands show a mixture of the two colors.
    \label{figbnds}}
\end{figure}

Finally, we consider the changes in band structure resulting from the
different structural arrangements. 
In Fig. \ref{figbnds} we show the band structure
of the rhombohedral perovskite compared to that of the monoclinic
structure along equivalent {\bf k}-directions. 
We can see that while in both cases
bonding-antibonding interactions between the Ge-$s$ and the Cl-$p$
orbitals determine the band edges of the valence band (VB), 
the antibonding bands become separated in the monoclinic case and 
the overall band width is significantly reduced. 
This reflects the disruption of the 3D corner sharing network
of Ge-$s$ and X-$p$ orbitals.  Also, the Ge-$p$ orbitals
which form the conduction band (CB) minimum become much flatter.
This is because in the highly symmetric perovskite
structure, these Ge-$p$ states at the Brillouin zone corner have no interactions
with other orbitals.\cite{Lingyi13} The lower symmetry of the monoclinic
structure allows more antibonding interactions between Ge-$p$ and I
raising this band at $R2$ and flattening out the lowest CB.
The splitting between the center of gravity
of the non-bonding X-$p$ bands and the Ge-$p$ remains more or less constant
because this is determined by the atomic energy levels. The net
result of the narrowing of the VB  and the flattening out of the
lowest CB, leads to a significant increase in the gap.
The flatter bands in the monoclinic case reflect the  molecular crystal like
structure of disconnected units compared to a continuous network. 

{\bf Conclusion:}
In summary, we have shown that the relative stability of the monoclinic
to the perovskite structure in AGeX$_3$  can be understood by
viewing the structure in terms of GeX$_3$ dipole carrying units.
The antiferroelectric arrangement of the dipoles in the monoclinic
structure is favored by electrostatics while the additional bond
formation for the ferroelectric arrangement stabilizes the
perovskite structure.  It is the competition between these two
effects which determines the crystal structure adapted. 
The fact that electrostatics plays a key role
suggests that the ferroelectric alignment
could be enforced by means of an applied electric field.   This would
be accompanied by  strong reduction in band band gap and other changes
in the electronic band structure.  The fact
that small changes in the dipole allowed us to
switch from one structure to the other and that intermediate types of
alignment of the GeX$_3$ units occur in these materials as function
of temperature suggests that this should be feasible. 

{\bf Methods} The calculations
were performed   using the all-electron
full-potential linearized muffin-tin orbital
method\cite{Methfessel,Kotani10} as implemented in the questaal-suite\cite{questaal}  and within the Perdew-Burke-Ernzerhof (PBE) \cite{PBE}
generalized gradient
approximation (GGA) to density functional theory (DFT).
These band structures are calculated  in 
the quasiparticle self-consistent (QS)$GW$ approximation\cite{MvSQSGWprl,Kotani07} where $G$ is the one-particle Greens' function and $W$
the screened Coulomb interaction (details in SM).

{\bf Acknowledgements:} This work was supported by the U.S. Department of Energy (Basic Energy Sciences)
  DOE-BES under grant No. DE-SC0008933. The calculations made use of the High Performance Computing Resource in the Core Facility for Advanced Research Computing at Case Western Reserve University.

\bibliography{Bib/csnx.bib,Bib/dft.bib,Bib/lmto.bib,Bib/gw.bib,Bib/spin}

\end{document}